\documentclass[twocolumn,aps,showpacs,preprintnumbers,amsmath,amssymb]{revtex4}

\usepackage{graphicx}
\usepackage{dcolumn}
\usepackage{bm}
\def\bra#1{\mathinner{\langle{#1}|}}
\def\ket#1{\mathinner{|{#1}\rangle}}

\begin{document}

\title{Measure for the Non-Markovianity of Quantum Processes}

\author{Elsi-Mari Laine$^{1}$\footnote{Electronic address:
emelai@utu.fi},
Jyrki Piilo$^{1}$\footnote{Electronic address: jyrki.piilo@utu.fi},
and Heinz-Peter Breuer$^{2}$\footnote{Electronic address:
breuer@physik.uni-freiburg.de} }
\affiliation{$^1$Turku Center for Quantum Physics, Department of Physics and Astronomy,
University of Turku, FI-20014, Turun Yliopisto, Finland \\
$^{2}$Physikalisches Institut, Universit\"at Freiburg,
Hermann-Herder-Strasse 3, D-79104 Freiburg, Germany}

\date{\today}

\begin{abstract}
Recently, a measure for the non-Markovian behavior of quantum
processes in open systems has been developed which is based on the
quantification of the flow of information between the open system
and its environment [Phys.~Rev.~Lett.~103, 210401 (2009)]. The
information flow is connected to the rate of change of the trace
distance between quantum states which can be interpreted in terms
of the distinguishability of these states. Here, we elaborate the
mathematical details of this theory, present applications to
specific physical models, and discuss further theoretical and
experimental implications, as well as relations to alternative
approaches proposed recently.
\end{abstract}


\maketitle

\section{Introduction}
A Markov process in the evolution of an open quantum system
typically gives rise to a quantum dynamical semigroup for which
the most general representation can be written in the Lindblad
form \cite{Gorini,Lindblad}. There exist however complex systems
for which this relatively simple description of the open system
dynamics in terms of a Markovian master equation fails to give a
comprehensive picture of the dynamics \cite{Breuer}. Thus in many
realistic physical systems the Markovian approximation of the
dynamics gives an overly simplified picture of the open system
evolution and a more rigorous treatment of the dynamics is
required.

To give insights into the nature of non-Markovian effects many
analytical methods and numerical simulation techniques have been
developed in recent years (see, for example, Refs.~\cite{nm
jumps,eisert,RIVAS,breuer2,breuer3,pascazio,krovi,lendi,budini,stenholm,wilkie,daffer,Kossakowski,shabani}).
Non-Markovianity manifests itself in the different approaches in a
variety of ways and there exists no general recipe for comparing
the degree of non-Markovianity in different physical models. In
order to give a general quantity determining the degree of
non-Markovian behavior in the open system dynamics, one has to
rigorously define what makes a dynamical map non-Markovian.

Here, we discuss a recently proposed measure for the degree of
non-Markovian behavior which is based on the trace distance
between quantum states \cite{nm-paper}. The trace distance
describes the probability of successfully distinguishing two
quantum states and the change in the trace distance of two open
system states can be interpreted as a flow of information between
the system and the environment. When the trace distance decreases
information flows from the system into the environment, while an
increase of the trace distance signifies a backflow of information
from the environment to the system. Markovian processes tend to
continuously decrease the distinguishability between any two
states of the open system, i.e., information flows continuously
from the system to the environment. The condition which defines a
non-Markovian dynamical map is that the map allows an information
flow from the environment to the system and, therefore, allows the
system to gain information about its former state. This condition
for a non-Markovian map leads to a rigorous and general definition
of a measure for the degree of non-Markovianity in open quantum
systems.

In Sec.~\ref{definition} we construct the measure for
non-Markovianity and discuss its properties for some general
classes of quantum processes in open systems. It is shown that the
non-divisibility of the dynamical map is necessary for the process
to be non-Markovian. Hence, the measure vanishes for quantum
dynamical semigroups and for time-dependent Markov processes. We
also demonstrate that the appearance of negative rates in the
quantum master equation is a necessary condition for
non-Markovianity. In Sec.~\ref{examples} we illustrate the
determination of the measure for a two-level system and for a
$\Lambda$-type atom in a cavity. Section \ref{discussion} contains
a detailed discussion of several alternative ways for defining a
measure for non-Markovianity. Moreover, we present possible
experimental strategies for the detection of non-Markovian
effects. The conclusions are drawn in Sec.~\ref{conclusions}.

\section{The Measure for Non-Markovianity}\label{definition}

\subsection{Construction of the measure}
To construct the measure for non-Markovianity we need a measure
for the distance between any pair of quantum states represented by
density matrices $\rho_1$ and $\rho_2$. Such a measure is given by
the trace distance, which is defined as
\begin{equation}
 D(\rho_1,\rho_2)=\frac{1}{2}\textrm{Tr}\left|\rho_1-\rho_2\right|,
 \label{eq:trace distance}
\end{equation}
where the modulus of an operator $A$ is defined by
$\left|A\right|=\sqrt{A^\dagger A}$. The trace distance $D$ yields
a natural metric on the state space and satisfies $0\leq D\leq 1$.
It has many nice properties that make it a useful measure for the
distance between quantum states \cite{NIELSEN}. First, the trace
distance is preserved under unitary transformations $U$,
\begin{equation}
    D(U\rho_1U^\dagger,U\rho_2U^\dagger)=D(\rho_1,\rho_2).
    \label{eq:unitary}
\end{equation}
Second, all completely positive and trace preserving (CPT) maps
$\Phi$ (trace preserving quantum operations) are contractions for
this metric,
\begin{equation}
    D(\Phi \rho_1,\Phi\rho_2)\leq D(\rho_1,\rho_2).
    \label{eq:contractivity}
\end{equation}
Third, the trace distance has a physical interpretation as a
measure of state distinguishability. Suppose Alice prepares a
quantum system in the state $\rho_1$ with probability $1/2$, and
in the state $\rho_2$ with probability $1/2$. She gives the system
to Bob, who performs a measurement to distinguish the two states.
The maximal probability that Bob can identify the state given to
him is \cite{GILCHRIST}
\begin{equation}
 p_{\max} = \frac{1}{2}\left[1 + D(\rho_1, \rho_2)\right].
\end{equation}
Hence, the trace distance represents the maximal bias in favor of
the correct state identification which Bob can achieve through an
optimal strategy. For example, if $\rho_1$ and $\rho_2$ have
orthogonal supports the trace distance becomes
$D(\rho_1,\rho_2)=1$ and thus $p_{\max}=1$, which means that Bob
is able to distinguish the states with certainty.

The change in the distinguishability of states of an open system
can be interpreted as a flow of information between the system and
the environment. We consider here quantum processes given by a
dynamical CPT map $\Phi(t,0)$ which transforms the initial states
$\rho(0)$ at time zero to the states $\rho(t)$ at time $t\geq 0$,
\begin{equation}
 \rho(0) \mapsto \rho(t) = \Phi(t,0)\rho(0).
\end{equation}
When such a quantum process reduces the distinguishability of
states, information is flowing from the system to the environment.
Likewise, the increase of the distinguishability signifies that
information flows from the environment to the system. The
invariance under unitary transformations \eqref{eq:unitary}
indicates that information is preserved under the dynamics of
closed systems. The contraction property of
Eq.~\eqref{eq:contractivity} guarantees that the maximal amount of
information the system can recover from the environment is the
amount of information earlier flowed out the system.

The basic idea underlying our construction for the measure of
non-Markovianity in a quantum process is that for Markovian
processes information flows continuously from the system to the
environment. In order to give rise to non-Markovian effects there
must be, for some interval of time, an information flow from the
environment back to the system. The information flowing from the
environment back to the system allows the earlier states of the
system to have an effect on the later dynamics of the system,
i.e., it allows the emergence of memory effects.

We define the rate of change of the trace distance of a pair of
states by means of
\begin{equation}
    \sigma(t,\rho_{1,2}(0))=\frac{d}{dt}D(\rho_1(t),\rho_2(t)),
    \label{eq:rate}
\end{equation}
where $\rho_{1,2}(t)=\Phi(t,0)\rho_{1,2}(0)$. For a non-Markovian
process described by a dynamical map $\Phi(t,0)$, information must
flow from the environment to the system for some interval of time
and thus we must have $\sigma>0$ for this time interval. A measure
of non-Markovianity should measure the total increase of
distinguishability over the whole time evolution, i.e., the total
amount of information flowing from the environment back to the
system. This suggests defining the measure $\mathcal{N}(\Phi)$ for
the non-Markovianity of the quantum process $\Phi(t,0)$ through
\begin{equation}
 \mathcal{N}(\Phi)=\max_{\rho_{1,2}(0)}\int_{\sigma>0}{dt\sigma(t,\rho_{1,2}(0))}.
 \label{eq:measure}
\end{equation}
The time integration is extended over all time intervals
$(a_i,b_i)$ in which $\sigma$ is positive and the maximum is taken
over all pairs of initial states. Due to Eq.~\eqref{eq:rate} the
measure can be written as
\begin{equation}
 \mathcal{N}(\Phi) =
 \max_{\rho_{1,2}(0)}\sum_i{\left[D(\rho_1(b_i),\rho_2(b_i))-D(\rho_1(a_i),\rho_2(a_i))\right]}.
 \label{eq:final measure}
\end{equation}
To calculate this quantity one first determines for any pair of
initial states the total growth of the trace distance over each
time interval $(a_i,b_i)$ and sums up the contribution of all
intervals. $\mathcal{N}(\Phi)$ is then obtained by determining the
maximum over all pairs of initial states. While it may be
difficult to derive an analytical expression for the measure
defined in Eq.~\eqref{eq:final measure}, the numerical evaluation
of the measure is relatively easy provided the dynamical map is
known explicitly. We will discuss in Sec.~\ref{examples} the
determination of $\mathcal{N}(\Phi)$ for some specific examples.

\subsection{Classification of quantum processes}
Having defined our measure for non-Markovianity we discuss in this
section the properties of this measure for some general classes of
quantum processes. Specific physical systems will be investigated
in Sec.~\ref{examples}.

\subsubsection{Divisible maps}\label{divisible maps}
The dynamical map $\Phi(t,0)$ is defined to be divisible if for
all $t,\tau\geq 0$ the CPT map $\Phi(t+\tau,0)$ can be written as
composition of the two CPT maps $\Phi(t+\tau,t)$ and $\Phi(t,0)$,
\begin{equation}
 \Phi(t+\tau,0)=\Phi(t+\tau,t)\Phi(t,0).
 \label{eq:divisibility}
\end{equation}
We note that this definition differs slightly from the usual
definition of divisibility according to which a CPT map $\Lambda$
(quantum channel) is said to be divisible if there exist CPT maps
$\Lambda_1$ and $\Lambda_2$ such that
$\Lambda=\Lambda_1\Lambda_2$, where it is assumed that neither
$\Lambda_1$ nor $\Lambda_2$ is a unitary transformation
\cite{div}. In Eq.~\eqref{eq:divisibility} the left-hand side as
well as the second factor on the right-hand side are fixed by the
given dynamical map. Hence, Eq.~\eqref{eq:divisibility} requires
the existence of a certain linear transformation $\Phi(t+\tau,t)$
which maps the states at time $t$ to the states at time $t+\tau$
and represents a CPT map (that may be a unitary transformation)
for all $t$ and all $\tau$. There are many quantum processes which
are not divisible. For instance, if $\Phi(t,0)$ is not invertible,
a linear map $\Phi(t+\tau,t)$ which fulfills
Eq.~\eqref{eq:divisibility} may not exist. Moreover, even if a
linear map $\Phi(t+\tau,t)$ satisfying Eq.~\eqref{eq:divisibility}
does exist, this map needs not be completely positive, and not
even positive.

We claim that all divisible dynamical maps are Markovian. To prove
this statement suppose that $\Phi(t,0)$ is divisible. For any pair
of initial states $\rho_{1,2}(0)$ we then have
\begin{equation}
 \rho_{1,2}(t+\tau)=\Phi(t+\tau,t)\rho_{1,2}(t).
 \label{eq:div for initial}
\end{equation}
Since $\Phi(t+\tau,t)$ is a CPT map we can apply the contraction
property \eqref{eq:contractivity} to obtain:
\begin{equation}
 D(\rho_1(t+\tau),\rho_2(t+\tau))\leq D(\rho_1(t),\rho_2(t)).
 \label{eq:reduction}
\end{equation}
This shows that for all divisible dynamical maps the trace
distance decreases monotonically, i.e.,
$\sigma(t,\rho_{1,2}(0))\leq 0$ and, therefore,
$\mathcal{N}(\Phi)=0$. Thus, we conclude that all divisible
processes are Markovian and that non-Markovian processes must
necessarily be described by a nondivisible dynamical map.

\subsubsection{Quantum dynamical semigroups}
The prototype of a Markovian dynamics is provided by a Markovian
master equation for the density matrix,
\begin{equation}
    \frac{d}{dt}\rho(t)=\mathcal{L}\rho(t),
    \label{eq:Markovian ME}
\end{equation}
with a generator in Lindblad form \cite{Gorini,Lindblad}
\begin{equation}
 \mathcal{L}\rho=-i[H,\rho]+\sum_i{\gamma_i\left[A_i\rho
 A_i^\dagger-\frac{1}{2}\left\{A_i^\dagger A_i,\rho\right\}
 \right]},   \label{eq:L-form}
\end{equation}
involving a time-independent Hamiltonian $H$ as well as
time-independent Lindblad operators $A_i$ and positive decay rates
$\gamma_i\geq 0$. Such a master equation leads to a dynamical
semigroup of CPT maps, $\Phi(t,0)=\exp(\mathcal{L}t)$. With
$\Phi(t+\tau,t)=\exp(\mathcal{L}\tau)$ the divisibility condition
\eqref{eq:divisibility} is trivially satisfied. Hence, we have
$\mathcal{N}(\Phi)=0$ for all dynamical semigroups, i.e., for all
processes described by a master equation in the Lindblad form.

\subsubsection{Time-dependent Markov processes}
The divisibility property holds for a much larger class of quantum
processes than those described by a master equation of the form
\eqref{eq:Markovian ME}. Suppose we have a time-local master
equation of the form
\begin{equation}
    \frac{d}{dt}\rho(t)=\mathcal{K}(t)\rho(t)
    \label{eq:tpMarkovian ME}
\end{equation}
with a time-dependent generator $\mathcal{K}(t)$. It can be shown
that in order to preserve the Hermiticity and trace of the density
matrix this generator must be of the form \cite{Gorini,breuer3}
\begin{eqnarray} \label{eq:tpL-form}
    \mathcal{K}(t)\rho&=&-i\left[H(t),\rho\right]\\
    &+&\sum_i{\gamma_i(t)\left[A_i(t)\rho A_i^\dagger(t)
    -\frac{1}{2}\left\{A_i^\dagger(t)A_i(t),\rho\right\} \right]}.
    \nonumber
\end{eqnarray}
By contrast to the assumptions in Eq.~\eqref{eq:L-form} the
Hamiltonian $H(t)$, the Lindblad operators $A_i(t)$ and the decay
rates $\gamma_i(t)$ may now depend on time. If the decay rates are
positive functions, $\gamma_i(t)\geq 0$, the generator
\eqref{eq:tpL-form} is in Lindblad form \eqref{eq:L-form} for each
fixed $t\geq 0$. Such a process with $\gamma_i(t)\geq 0$ may be
called time-dependent Markovian although the corresponding
dynamical map
\begin{equation}
 \Phi(t,0) = {\rm T} \exp\left[\int_{0}^{t}{dt'\mathcal{K}(t')}\right]
 \label{eq:tp CPT}
\end{equation}
does not yield a dynamical semigroup (${\rm T}$ denotes the
chronological time-ordering operator). However, one can easily see
that the divisibility condition \eqref{eq:divisibility} still
holds because the map
\begin{equation}
 \Phi(t+\tau,t) = {\rm T} \exp\left[\int_t^{t+\tau}{dt'\mathcal{K}(t')}\right]
 \label{eq:div for tp Mark}
\end{equation}
is CPT for $\gamma_i(t)\geq 0$. Thus we can conclude that for all
time-dependent Markovian processes we again have
$\mathcal{N}(\Phi)=0$.

We have just seen that a quantum process given by the time-local
master equation \eqref{eq:tpMarkovian ME} with positive rates
leads to a divisible dynamical map. Under certain conditions the
converse of this statement is also true. More precisely, if the
dynamical map $\Phi(t,0)$ is divisible with a unique map
$\Phi(t+\tau,t)$ depending smoothly on $\tau$, then the
corresponding density matrix $\rho(t)$ obeys a master equation of
the form \eqref{eq:tpMarkovian ME} with positive rates in the
generator \eqref{eq:tpL-form}. In fact, using
$\rho(t+\tau)=\Phi(t+\tau,t)\rho(t)$ we find
\begin{equation}
 \frac{d}{dt}\rho(t) =
 \left.\frac{d}{d\tau}\right|_{\tau=0} \Phi(t+\tau,t)\rho(t),
\end{equation}
and, hence, we obtain the master equation \eqref{eq:tpMarkovian
ME}, where the generator is given by
\begin{equation}
 {\mathcal{K}}(t) = \left.\frac{d}{d\tau}\right|_{\tau=0}
 \Phi(t+\tau,t).
\end{equation}
Since $\Phi(t+\tau,t)$ is CPT and satisfies $\Phi(t,t)=I$, this
generator must be in Lindblad form for each fixed $t$, i.e., it
must have the form \eqref{eq:tpL-form} with $\gamma_i(t)\geq 0$.

\subsubsection{Non-Markovian processes}
The measure for quantum non-Markovianity does not depend on any
specific mathematical representation of the dynamics. There are
many different such representations, e.g., through generalized
master equations involving a certain memory kernel. However,
quantum master equations with the time-local structure given by
Eqs.~\eqref{eq:tpMarkovian ME} and \eqref{eq:tpL-form} are also
very useful for the description of non-Markovian processes. It
follows from the preceding results that in order for such a master
equation to yield a nonzero measure, $\mathcal{N}(\Phi)>0$, at
least one of the rates $\gamma_i(t)$ must take on negative values
for some interval of time. We emphasize that temporarily negative
rates in the master equation do in general not lead to a violation
of the complete positivity of the dynamical map. Many examples for
time-local master equations with negative rates are known in the
literature. Further examples will be discussed in the next
section.

\section{Examples}\label{examples}

\subsection{Two-level system}\label{tl}
We study the dynamics of a two-level atom with excited state
$\ket{+}$ and ground state $\ket{-}$ which is coupled to a
reservoir of field modes initially in the vacuum state. In
Ref.~\cite{nm-paper} we have described the detuned Jaynes-Cummings
model, while here we treat the resonant case. We will show that
the pair of states maximizing the measure for non-Markovinity is
different in the two cases. This demonstrates that the change in
both the populations and the coherences plays a crucial role in
the flow of information between the system and the environment.

The two-level atom model
can easily be solved exactly \cite{Breuer} and leads to a
dynamical map $\Phi(t,0)$ which can be represented in terms of the
elements $\rho_{\pm\pm}(t)=\langle \pm|\rho(t)|\pm\rangle$ of the
density matrix $\rho(t)$ as follows,
\begin{eqnarray}
 \rho_{++}(t) &=& |G(t)|^2 \rho_{++}(0), \nonumber \\
 \rho_{--}(t) &=& \rho_{--}(0) + (1-|G(t)|^2)\rho_{++}(0), \nonumber \\
 \rho_{+-}(t) &=& G(t) \rho_{+-}(0), \nonumber \\
 \rho_{-+}(t) &=& G^*(t) \rho_{-+}(0). \label{DYN-MAP-1}
\end{eqnarray}
Here, the function $G(t)$ is defined as the solution of the
integrodifferential equation
\begin{equation} \label{G-DEF}
 \frac{d}{dt}G(t) = -\int_0^t dt_1 f(t-t_1)G(t_1)
\end{equation}
corresponding to the initial condition $G(0)=1$, where $f(t-t_1)$
denotes the two-point reservoir correlation function (Fourier
transform of the spectral density). The map \eqref{DYN-MAP-1} is
completely positive if and only if $|G(t)|\leq 1$. One can easily
check that $\Phi(t,0)$ can be decomposed as in
Eq.~\eqref{eq:divisibility}, where the map $\Phi(t+\tau,t)$ is
given by
\begin{eqnarray}
 \rho_{++}(t+\tau) &=& \left|\frac{G(t+\tau)}{G(t)}\right|^2 \rho_{++}(t), \nonumber \\
 \rho_{--}(t+\tau) &=& \rho_{--}(t)
 + \left(1-\left|\frac{G(t+\tau)}{G(t)}\right|^2\right)\rho_{++}(t), \nonumber \\
 \rho_{+-}(t+\tau) &=& \frac{G(t+\tau)}{G(t)} \rho_{+-}(t), \nonumber \\
 \rho_{-+}(t+\tau) &=& \frac{G^*(t+\tau)}{G^*(t)} \rho_{-+}(t). \label{DYN-MAP-2}
\end{eqnarray}
It follows from these equations that a necessary and sufficient
condition for the complete positivity of $\Phi(t+\tau,t)$ is given
by $|G(t+\tau)|\leq |G(t)|$. Thus we see that the dynamical map of
the model is divisible if and only if $|G(t)|$ is a monotonically
decreasing function of time. Note that this statement holds true
also for the case that $G(t)$ vanishes at some finite time.

With the help of the above results one can easily derive an
analytical formula for the time derivative of the trace distance,
\begin{equation}
 \sigma(t,\rho_{1,2}(0)) = \frac{2|G(t)|^2a^2 + |b|^2}{\sqrt{|G(t)|^2a^2 + |b|^2}}
 \frac{d}{dt}|G(t)|,
 \label{eq:sigma for two-level}
\end{equation}
where $a=\rho_1^{++}(0)-\rho_2^{++}(0)$ denotes the difference of
the populations and $b=\rho_1^{+-}(0)-\rho_2^{+-}(0)$ the
difference of the coherences of the initial states. This relation
shows that the trace distance increases at some point if and only
if $|G(t)|$ increases at this point. We conclude that the measure
for non-Markovianity is positive, $\mathcal{N}(\Phi)>0$, if and
only if the dynamical map is nondivisible.

A positive measure for non-Markovianity is not only linked to a
breakdown of the divisibility of the dynamical map, but also to
the emergence of a negative rate in the corresponding master
equation \eqref{eq:tpMarkovian ME}. In fact, as long as $G(t)\neq
0$ one can write an exact master equation of this form with the
generator
\begin{eqnarray}
\label{TCL-GEN}
 \mathcal{K}(t)\rho &=& -\frac{i}{2}S(t)
 [\sigma_+\sigma_-,\rho] \\
 &~& +\gamma(t)\left[ \sigma_-\rho\sigma_+
 -\frac{1}{2}\left\{\sigma_+\sigma_-,\rho\right\} \right], \nonumber
\end{eqnarray}
where we use the definitions
\begin{equation}
 \gamma(t) = -2\Re\left(\frac{\dot{G}(t)}{G(t)}\right), \qquad
 S(t) = -2\Im\left(\frac{\dot{G}(t)}{G(t)}\right).
\end{equation}
Writing the rate $\gamma(t)$ as
\begin{equation}
 \gamma(t) = -\frac{2}{|G(t)|}\frac{d}{dt}|G(t)|
\end{equation}
we see that an increase of $|G(t)|$ and, hence, a breakdown of the
divisibility leads to a negative rate in the generator of the
master equation. Thus we find that for the present model a nonzero
measure for non-Markovianity is equivalent to the non-divisibility
of the dynamical map and to the occurrence of a negative rate in
the master equation.

\begin{figure}[htb]
\includegraphics[width=0.35\textwidth]{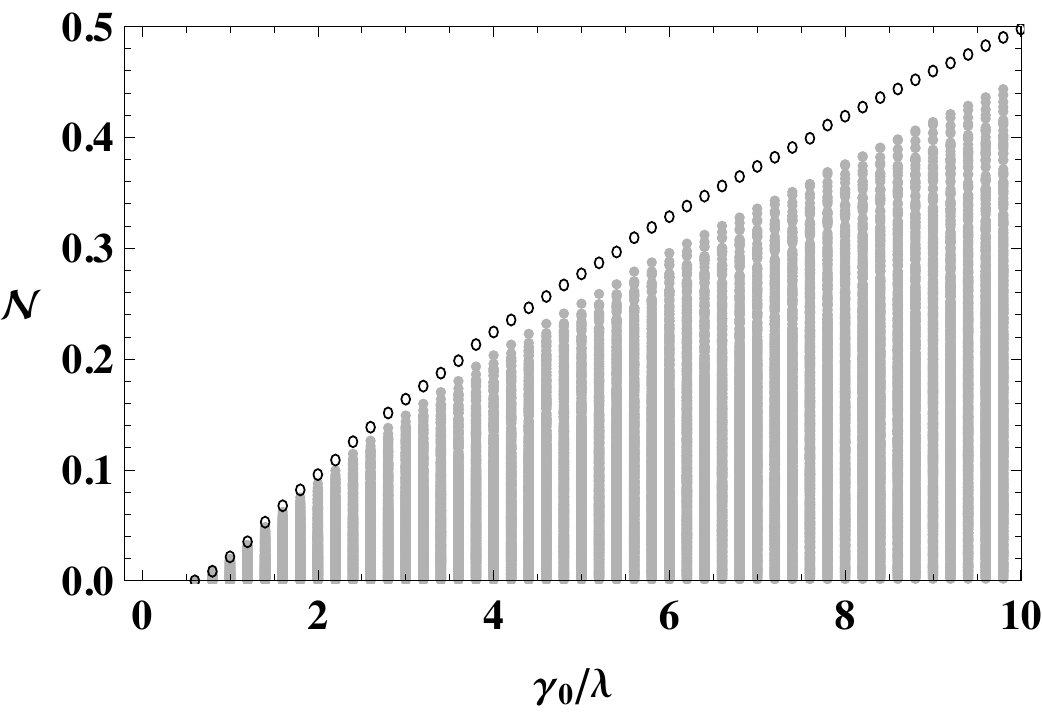}
\caption{The non-Markovianity $\mathcal{N}(\Phi)$ for the damped
Jaynes-Cummings model as a function of the coupling strength
$\gamma_0$. Gray dots: 1000 randomly drawn pairs of pure and mixed
initial states. Black circles: The initial pair given by
Eq.~\eqref{INIT} which leads to the maximum in Eq.
\eqref{eq:measure}.} \label{twolevelmax}
\end{figure}

As an example we consider the case of a Lorentzian reservoir
spectral density which is on resonance with the atomic transition
frequency and leads to an exponential two-point correlation
function
\begin{equation} \label{exp-corr}
  f(\tau) = \frac{1}{2} \gamma_0 \lambda e^{-\lambda |\tau|},
\end{equation}
where $\gamma_0$ describes the coupling strength and $\lambda$ the
spectral width (damped Jaynes-Cummings model). Solving
Eq.~\eqref{G-DEF} with this correlation function we find
\begin{equation}
 G(t) = e^{-\lambda t/2}\left[\cosh\left(\frac{dt}{2}\right)
 +\frac{\lambda}{d}\sinh\left(\frac{dt}{2} \right)\right],
 \label{eq:GT}
\end{equation}
where $d=\sqrt{\lambda^2-2\gamma_0\lambda}$. We see that for small
couplings, $\gamma_0<\lambda/2$, the function $|G(t)|$ decreases
monotonically. The dynamical map is thus divisible in the weak
coupling regime, the rate $\gamma(t)$ is positive, and the measure
for non-Markovianity vanishes. However, in the strong coupling
regime, $\gamma_0>\lambda/2$, the function $|G(t)|$ starts to
oscillate, showing a non-monotonic behavior. Consequently, the
dynamical map is then no longer divisible and
$\mathcal{N}(\Phi)>0$. We note that in the strong coupling regime
the rate $\gamma(t)$ diverges at the zeros of $G(t)$. However, the
master equation can still be used to describe the evolution
between successive zeros and, therefore, the connection between a
positive measure and negative rates in the master equation remains
valid.

There is thus a threshold $\gamma_0 = \lambda/2$ for the
system-reservoir coupling below which $\mathcal{N}(\Phi)=0$. We
find that the measure increases monotonically with increasing
coupling for $\gamma_0>\lambda/2$. This is illustrated in
Fig.~\ref{twolevelmax}. The maximization over the pair of initial
states $\rho_{1,2}(0)$ in expression \eqref{eq:measure} has been
performed here by a Monte Carlo sampling of pairs of initial
states. Our simulations provide strong evidence that the maximum
is attained for the initial states
\begin{equation} \label{INIT}
 \rho_1(0)=\ket{-}\bra{-}, \;\;\;
 \rho_2(0)=\frac{1}{2}(\ket{+}+\ket{-})(\bra{+}+\bra{-}).
\end{equation}
In Ref.~\cite{nm-paper} we calculated the measure for the detuned
Jaynes-Cummings model in the weak coupling limit. In this example
the maximum of the measure was obtained for the initial states
$\rho_1(0)=\ket{-}\bra{-}$ and $\rho_2(0)=\ket{+}\bra{+}$, i.e.
for the invariant ground state and the excited state. The
difference in the maximization for the resonant and the
off-resonant case arises from the fact that the rate at which the
populations and the coherences initially decay is much larger for
the resonant case. Consequently, the growth of the trace distance
occurs after the excited state population and the coherences have
reached the value zero. After this point, the increase of the
coherences yields the dominant contribution to the increase of the
trace distance. Therefore, the maximal growth of the trace
distance for the resonant case is reached for the invariant state
and the state with maximal initial coherence.

\subsection{$\Lambda$-model} \label{LAMBDA-MODEL}
The $\Lambda$-model describes a three-level atom with excited
state $\ket{a}$ and two ground states $\ket{b}$ and $\ket{c}$
interacting off-resonantly with a cavity field. This example allows us
to demonstrate how the measure for non-Markovianity  operates
in a multi-channel case and how there can exist simultaneously
positive and negative decay rates for different channels.
The spectral
density we use is
\begin{equation}
 J(\omega) = \frac{\gamma_0}{2\pi}\frac{\lambda^2}{(\omega_{\rm cav}-\omega)^2+\lambda^2},
\end{equation}
where $\omega_{\rm cav}$ is the resonance frequency of the cavity.
Further details and the master equation describing the dynamics of
the $\Lambda$-type atom are presented in the Appendix. The
generator of the master equation is of the form of
Eq.~\eqref{eq:tpL-form} with two Lindblad operators
$\ket{b}\bra{a}$ and $\ket{c}\bra{a}$, and two time-dependent
decay rates $\gamma_1(t)$ and $\gamma_2(t)$.

\begin{figure}[htb]
\includegraphics[width=0.35\textwidth]{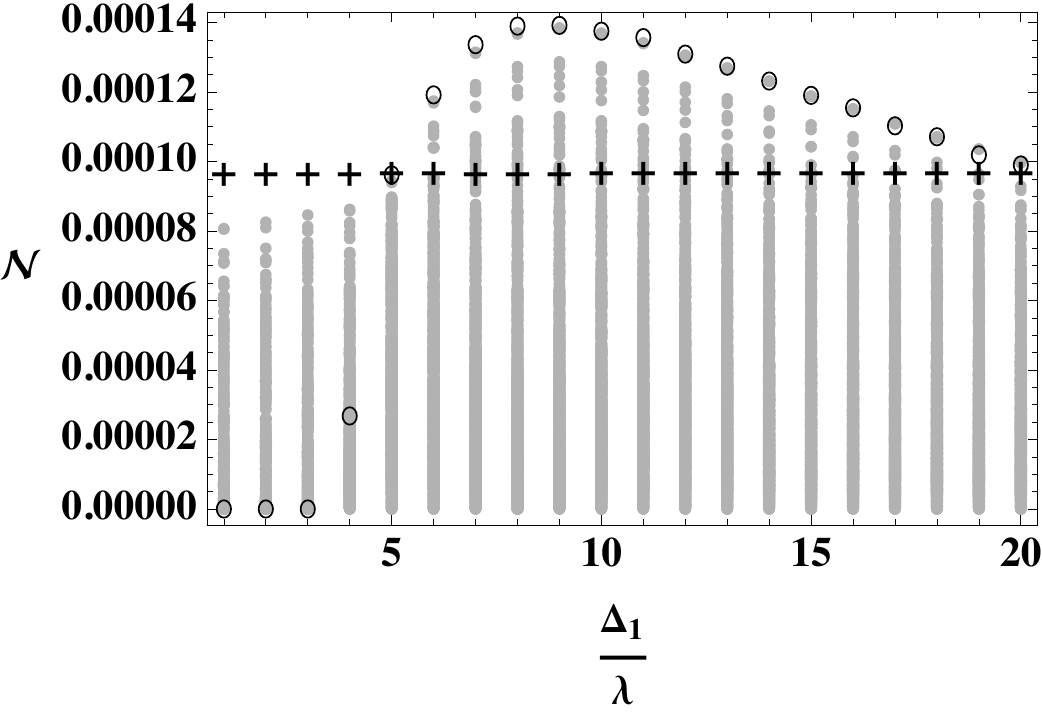}
\caption{The non-Markovianity $\mathcal{N}(\Phi)$ for the
$\Lambda$-model as a function of the detuning $\Delta_1$ for
$\Delta_2/\lambda=5$ and $\gamma_0/\lambda=0.01$. Gray dots: 1000
randomly drawn pairs of initial states. Circles: The initial pair
$\rho_1(0)=\ket{a}\bra{a}$ and $\rho_2(0)=\ket{b}\bra{b}$. Pluses:
The initial pair $\rho_1(0)=\ket{a}\bra{a}$ and
$\rho_2(0)=\ket{c}\bra{c}$. At $\Delta_1=\Delta_2$ the pair which
yields the maximum in Eq.~(\ref{eq:measure}) changes from the
latter to the former pair of initial states.} \label{lambda}
\end{figure}

The detunings of the transition frequencies of the $\Lambda$-atom
from the cavity resonance frequency are denoted by
$\Delta_i=\omega_i-\omega_\textrm{cav}$. When the detuning
parameters $\Delta_1$ and $\Delta_2$ are both sufficiently large,
the decay rates $\gamma_1(t)$ and $\gamma_2(t)$ get temporarily
negative values and this gives rise to an information flow  from
the environment to the system. On one hand, the two decay rates
$\gamma_1(t)$ and $\gamma_2(t)$ have simultaneous negative regions
when $\Delta_1=\Delta_2$. On the other hand, when
$\Delta_1\neq\Delta_2$, the decay rates can have opposite signs.
In this case, the co-operative action of the other channel reduces
the amount of information flowing from the environment to the
system. The maximum of the measure over the initial states is
reached when the states are chosen to be $\ket{a}\bra{a}$ and
$\ket{b}\bra{b}$, or $\ket{a}\bra{a}$ and $\ket{c}\bra{c}$,
depending on which of the channels has more information flow from
the environment to the system. When $\Delta_1$ and $\Delta_2$ are
such that the channel corresponding to the decay rate
$\gamma_{i}(t)$ ($i=1$ or $2$) causes more information flow from
the environment to the system we get the expression
\begin{equation}
 \sigma(t) = -\gamma_{i}(t)\rho_{aa}(t).
 \label{eq:sigma for lambda}
\end{equation}
The function $\rho_{aa}(t)$ is specified in the Appendix.
Eq.~\eqref{eq:sigma for lambda} shows that the $\Lambda$-system is
non-Markovian if one of the decay rates $\gamma_1(t)$ or
$\gamma_2(t)$ takes on negative values. The maximization over the
the pair of initial states is demonstrated in Fig. \ref{lambda},
where the measure was again calculated numerically from a large
sample of initial states.

\section{Discussion}\label{discussion}

\subsection{Alternative distance measures}
We have based our definition of the measure of non-Markovianity on
the trace distance \eqref{eq:trace distance}. An alternative
measure is obtained if one replaces the trace distance by the
relative entropy
\begin{equation}
 S(\rho_1||\rho_2)=\textrm{Tr}\left[\rho_1(\log\rho_1-\log\rho_2)\right].
 \label{eq:relative entropy}
\end{equation}
Using this quantity as a measure for the distance between quantum
states one is led to a similar interpretation as before because
the relative entropy also decreases under CPT maps
\cite{lindblad2}. There are however some technical problems and
limitations in the usefulness of the relative entropy which arise
from the fact that for many pairs $\rho_1$ and $\rho_2$ the
relative entropy becomes infinite \cite{lendi2} and thus leads to
singularities in the definition of the measure. This situation can
occur even in the simple case of a two-state system, demonstrating
the problems of the relative entropy concept in defining a general
measure for non-Markovianity. No such problems occur for the trace
distance which is well-defined and finite for all physical states
represented by positive trace class operators.

Another common measure for the distance between two states is the
Hilbert-Schmidt distance
\begin{equation}
    D_\textrm{HS}(\rho_1,\rho_2)=\sqrt{\textrm{Tr}\left[(\rho_1-\rho_2)^2\right]}.
    \label{eq:Hilbert-Schmidt}
\end{equation}
For two-dimensional Hilbert spaces the Hilbert-Schmidt distance
and the trace distance coincide and correspond to the Euclidean
distance between the Bloch vectors representing the states (up to
numerical factors). However, the Hilbert-Schmidt distance is not
suitable for a definition of non-Markovianity since CPT maps are
in general not contractions for this metric \cite{wang}. Thus, the
Hilbert-Schmidt distance does not provide a natural way to define
the information flow between system and environment.

\subsection{Experimental issues}
The exact determination of the measure generally requires solving
the complete reduced dynamics which can be a difficult task for
more complex systems. However, any observed growth of the trace
distance is a clear signature for non-Markovian behavior and leads
to a lower bound for $\mathcal{N}(\Phi)$. The measure for
non-Markovianity introduced here could therefore be useful also
for the experimental detection of non-Markovianity.

In an experiment one has to perform a state tomography on
different ensembles at different times in order to decide whether
or not the trace distance has increased. Such an experiment also
allows the validation of theoretical models or approximation
schemes. Consider a theoretical model predicting
$\sigma(t,\rho_{1,2}(0))>0$ for some interval $t\in(t_1,t_2)$ and
for some pair of initial states $\rho_{1,2}(0)$. In the experiment
one should then detect the increase of the trace distance between
the states $\rho_1(t)$ and $\rho_2(t)$ in this time interval. This
type of experiment could be based, e.~g., on the recent proposal
to use a trapped ion to study quantum Brownian motion in the
non-Markovian regime \cite{QBM}. The explicit experimental
implementation of this system can be done, e.~g., by using
reservoir engineering techniques \cite{Wineland} or by using the
trapped ion as a quantum simulator for non-Markovian dynamics
\cite{Qsimu}. One of the possibilities here to detect
non-Markovianity is to prepare the ion in various Fock states, and
to study the trace distance dynamics as described above.

A great advantage of the present approach is that it also allows
to plan experiments for testing non-Markovianity without knowing
the properties of the environment or the system-environment
interaction. The interactions and environmental properties can be
quite difficult to model in an experimental setup. By performing a
state tomography for two states of the open system under study at
many different times, one can determine whether there has been any
increase in the trace distance and, hence, non-Markovian behavior
in the dynamics. From this information one can conclude whether or
not non-Markovian effects are crucial in the dynamics and in this
way also gain some knowledge on the nature of the environment and
the interactions. An example under active investigation, where
nevertheless a complete characterization of the environment is
still missing and where non-Markovianity could play a role, is
given by the energy transfer in photosynthetic systems
\cite{Engel}.

\subsection{Other approaches to non-Markovianity}
Recently, other interesting approaches to the characterization and
quantification of non-Markovianity have been proposed. The measure
suggested in Ref.~\cite{eisert} quantifies non-Markovianity in
terms of the minimal amount of noise required to make a given
quantum channel Markovian. The most important difference to our
approach is that this measure is based on the properties of the
dynamical map at a given time, i.e., on the properties of the
quantum channel represented by a snapshot of the time evolution.
Hence, this approach assesses to what extend the dynamical map at
each fixed time $t_0$ deviates from an element of a Markovian
process. The fundamental difference between the notion of
non-Markovianity used in Ref.~\cite{eisert} and ours can be seen
from the following simple example. We consider the dynamical map
$\Phi(t,0)$ of a two state system undergoing a pure de- and
re-phasing dynamics which is given by (using the notation of
Sec.~\ref{tl})
\begin{eqnarray}
 \rho_{++}(t) &=& \rho_{++}(0), \quad
 \rho_{--}(t) = \rho_{--}(0), \nonumber \\
 \rho_{+-}(t) &=& g(t) \rho_{+-}(0), \quad
 \rho_{-+}(t) = g(t) \rho_{-+}(0), \label{DEPHASING}
\end{eqnarray}
where the function $g(t) = \frac{1}{2}[1+\cos^2\omega t]$
describes a periodic oscillation of the coherences. The trace
distance for this model is given by
\begin{equation}
 D(\rho_1(t),\rho_2(t)) = \sqrt{a^2+g^2(t)|b|^2}
\end{equation}
where $a=\rho_1^{++}(0)-\rho_2^{++}(0)$ and
$b=\rho_1^{+-}(0)-\rho_2^{+-}(0)$. For $b\neq 0$ the trace
distance thus oscillates periodically and, hence,
$\mathcal{N}(\Phi)=+\infty$ according to the definition
\eqref{eq:measure} of our measure. On the other hand, the
non-Markovianity in the sense of Ref.~\cite{eisert} is zero
because for any fixed $t_0$ the dynamical map \eqref{DEPHASING}
can be written as an element of a Markovian semigroup:
$\Phi(t_0,0)=\exp({\mathcal{L}})$ with the Lindblad generator
${\mathcal{L}}\rho=\frac{\Gamma}{2}(\sigma_3\rho\sigma_3-\rho)$,
where $\Gamma=-\ln g(t_0)$.

A further interesting measure proposed recently \cite{RIVAS} is
closely connected to the measure discussed here. In fact, the
measure of Ref.~\cite{RIVAS} quantifies deviations from the
divisibility of the dynamical map. As we have seen, the
non-divisibility of the dynamical map is a necessary condition for
$\mathcal{N}(\Phi)$ to be nonzero. However, we conjecture that our
notion of non-Markovianity and the one used in \cite{RIVAS} are
not strictly equivalent, i.e., that there are nondivisible maps
with $\mathcal{N}(\Phi)=0$. Further considerations concerning this
point will be published elsewhere.

\section{Conclusions}\label{conclusions}
We have constructed a measure $\mathcal{N}(\Phi)$ for the
non-Markovianity of quantum processes in open systems in terms of
the information flowing from the environment to the system during
the time evolution. The flow of information is characterized by
the change of the distinguishability between a pair of quantum
states which, in turn, is linked to the change of the trace
distance between these states. We have also argued why the trace
distance represents the most suitable distance measure for quantum
states to be used in this context. Furthermore, since we have
developed a genuine quantitative measure, the results presented
here also allow to compare the degree of non-Markovianity of
different types of physical systems.

It has been demonstrated that a nonzero measure for
non-Markovianity requires the dynamical map to be nondivisible, a
property which is thus necessary for the presence of memory
effects in the open system dynamics. It has also been shown that
Markovian semigroups and time-dependent Markov processes are
divisible and, hence, lead to $\mathcal{N}(\Phi)=0$. The examples
discussed here illustrate how the measure can be calculated for a
given open system dynamics and that a nonzero measure for
non-Markovianity is linked to the emergence of negative decay
rates in the corresponding master equation.

Our measure for non-Markovianity has a clear operational meaning
based on the interpretation of the trace distance in terms of the
distinguishability of states, and suggests various ways to
experimentally decide whether a system under study is
non-Markovian. The measurement scheme discussed here has the great
advantage that it does not presuppose any knowledge about the
structure of the environment or about the system-environment
interaction and, therefore, also gives valuable information on the
theoretical modelling of the open system dynamics. If, for
example, a substantial increase of the trace distance is observed
experimentally, a mathematical description of the dynamics through
any equation describing a Markovian or time-dependent Markovian
process is excluded. This shows that our measure is a useful tool
for the characterization of non-Markovianity, both in experiments
on open systems and in their theoretical analysis and modelling.

We have argued that the characteristics of the information
exchange between the system and its environment determine the
degree of non-Markovian behavior in an open system. This exchange
of quantum information has been defined here in very general terms
through the change of the distinguishability of quantum states,
and does not presuppose anything about the specific physical
carriers of the information, e.g., energy or particles. Moreover,
the measure does not depend on any specific representation of the
open system's dynamics. It therefore opens the possibility to
compare and assess different mathematical formulations of
dynamical processes in their ability to describe memory effects,
in order to understand better the mathematical description of
non-Markovian quantum dynamics.

\begin{acknowledgments}
The authors thank the Academy of Finland (projects 133682, 115982,
115682), Magnus Ehrnrooth Foundation, and the Finnish National
Graduate School of Modern Optics and Photonics for financial
support.
\end{acknowledgments}

\appendix*

\section{}

Here we present some details of the $\Lambda$-model studied in
Sec.~\ref{LAMBDA-MODEL}. The weak-coupling master equation for
this model is given by
\begin{eqnarray*}
\frac{d}{dt}\rho(t)&=&-i\lambda_1(t)\left[\ket{a}\bra{a},\rho(t)\right]-i\lambda_2(t)\left[\ket{a}\bra{a},\rho(t)\right]\nonumber\\
    &&+\gamma_1(t)\left[\ket{b}\bra{a}\rho(t)\ket{a}\bra{b}-\frac{1}{2}\left\{\rho(t),\ket{a}\bra{a}\right\} \right]\nonumber\\
    &&+\gamma_2(t)\left[\ket{c}\bra{a}\rho(t)\ket{a}\bra{c}-\frac{1}{2}\left\{\rho(t),\ket{a}\bra{a}\right\} \right],\nonumber
\end{eqnarray*}
where
\begin{eqnarray*}
    \lambda_i(t)&=&\int_0^t{ds\int_0^\infty{d\omega}J(\omega)\sin\left[(\omega-\omega_i)s\right]},\\
    \gamma_i(t)&=&\int_0^t{ds\int_0^\infty{d\omega}J(\omega)\cos\left[(\omega-\omega_i)s\right]}.
    \label{eq:decrates}
\end{eqnarray*}
Introducing the definitions
\begin{eqnarray*}
 f(t) &=& e^{-\left[D_1(t)+D_2(t)\right]/2}e^{-i\left[L_1(t)+L_2(t)\right]}, \\
 g_i(t)&=&\int_0^t{ds \gamma_i(s)e^{-[D_1(s)+D_2(s)]}},
 \label{eq:notations}
\end{eqnarray*}
where
\[
 D_i(t) = \int_0^t{ds\gamma_i(s)}, \qquad
 L_i(t) = \int_0^t{ds\lambda_i(s)},
\]
the solution of the master equation can be represented as follows,
\begin{eqnarray*}
 \rho_{aa}(t) &=& |f(t)|^2 \rho_{aa}(0),\\
 \rho_{bb}(t) &=& g_1(t) \rho_{aa}(0) + \rho_{bb}(0),\\
 \rho_{cc}(t) &=& g_2(t) \rho_{aa}(0) + \rho_{cc}(0),\\
 \rho_{ab}(t) &=& f(t) \rho_{ab}(0),\\
 \rho_{ac}(t) &=& f(t) \rho_{ac}(0),\\
 \rho_{bc}(t) &=& \rho_{bc}(0).
\end{eqnarray*}
These equations define the dynamical map $\Phi(t,0)$ of the
$\Lambda$-model. Employing the results of Choi \cite{Choi} one can
check that a necessary and sufficient condition for the complete
positivity of this map is given by
\[
 g_1(t) \geq 0, \quad g_2(t)\geq 0.
 \label{eq:conditions}
\]
These conditions are satisfied for the parameters used in the
simulations of Sec.~\ref{LAMBDA-MODEL}.

\end{document}